\documentclass[prl,showpacs,preprintnumbers,amsmath,citeautoscript, article, twocolumn]{revtex4}
\usepackage{graphicx}
\usepackage{dcolumn}
\usepackage{bm}
\usepackage{natbib}

\begin{document}


\title{Pyramidal micro-mirrors for microsystems and atom chips}

\author{M. Trupke}
\email{michael.trupke@imperial.ac.uk\\}
\author{F. Ramirez-Martinez}
\author{E.A. Curtis}
\author{J.P. Ashmore}
\author{S. Eriksson}
\author{E.A. Hinds} \affiliation{Blackett Laboratory, Imperial College,
Prince Consort Road, London SW7 2BW, United Kingdom}

\author{Z. Moktadir}
\author{C. Gollasch}
\author{M. Kraft}
\affiliation{School of Electronics and Computer Science,
Southampton University, Southampton, SO17 1BJ, United Kingdom}

\author{G. Vijaya Prakash}
\author{J.J. Baumberg}
\affiliation{School of Physics and Astronomy, Southampton
University, Southampton, SO17 1BJ, United Kingdom}


\begin{abstract}
Concave pyramids are created in the $(100)$ surface of a silicon
wafer by anisotropic etching in potassium hydroxide. High quality
micro-mirrors are then formed by sputtering gold onto the smooth
silicon (111) faces of the pyramids. These mirrors show great
promise as high quality optical devices suitable for integration
into MOEMS and atom chips. We have shown that structures of this
shape can be used to laser-cool and hold atoms in a
magneto-optical trap.
\end{abstract}

\maketitle

The miniaturization of optical components leads to
 higher packaging density and increased speed of devices that manipulate
light.  This is part of the vast field of Microsystems technology,
designated by Micro-Opto-Electro-Mechanical Systems (MOEMS), in
which electronic, mechanical, and optical devices are integrated
on the micron scale. As mirrors are fundamental components of most
optical systems, techniques for the integration of high-quality
mirrors are relevant for the advancement of this field. In the
context of atomic physics, there has been a recent drive to
integrate optical elements with atom
chips\cite{{chipReview},{eriksson},{cavities}}  for the purposes
of detection and quantum-coherent manipulation of cold
atoms\cite{{ScheelQIP},{singAtDet}}. Just as pyramidal mirrors
have been used\cite{pyrMOT} to form macroscopic magneto-optical
traps (MOTs), so these microscopic pyramids may be used to cool
and trap an array of small atom clouds on a chip.

\begin{figure}[!htb]
\begin{center}
\begin{tabular}{cc}
\includegraphics[width=8.5cm]{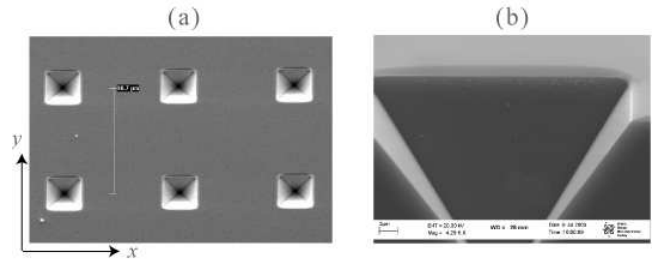}
\end{tabular}
\caption{SEM micrographs of the etched and gold-coated pyramids.
(a) top view showing pyramids in a rectangular array with a pitch
of $100\,\mu$m. (b) Cross-sectional view of a single pyramid. This
was obtained by cleaving the pyramid parallel to one of its edges.
The base of the pyramid has a side of length $30\,\mu$m,
corresponding to a perpendicular depth of $21.3\,\mu$m
}\label{SEM}
\end{center}
\end{figure}

We have fabricated 2-dimensional arrays of micro-mirrors in
silicon using a method that is simple, economical, and compatible
with MOEMS. We start with a $(100)$-oriented silicon wafer, coated
with a thin layer of oxide. Optical lithography is then used to
make square openings in the oxide, through which the silicon can
be etched. We use the anisotropic etchant potassium hydroxide
(KOH) at a concentration of $25\%$ by volume and a temperature of
$80\,{^\circ}$C. This attacks the Si$(100)$ plane more rapidly
than the Si$(111)$ plane, resulting in a pyramidal
pit~\cite{Brendel} bounded by the four surfaces $(1,1,1)$,
$(\bar{1},1,1)$, $(1,\bar{1},1)$ and $(\bar{1},\bar{1},1)$.
Typical resulting pyramids are shown in Fig.\,\ref{SEM}. The
Si$(111)$- faces of the pyramids are expected to be extremely
smooth because of the
 layer-by-layer etching mechanism involved\cite{{Moktadir},{Sato2}}. Atomic force
microscope measurements confirm this, giving an rms surface
roughness value of less than $0.5\,$nm
 for the uncoated pyramid faces. This makes them ideal as substrates for high-quality optical mirrors.
 After stripping the oxide mask away, a layer of gold of 100
nanometers thickness is applied to the silicon. Gold was chosen as
it is a good reflector for infrared light, but other metals or
dielectric coatings can also be applied. After sputtering gold,
the surface roughness increases to $3\,$nm (rms).
 With this amount of roughness one can calculate that the scattering loss
of the specularly reflected intensity should be less than  $0.5\%$
in the near-infrared range~\cite{Beckmann}.

Fig.\,\ref{SEM}\,(a) shows a small section of the array viewed
under a scanning electron microscope after completion of the gold
coating. In this particular sample, the square pyramids have
$30\,\mu$m sides and are arranged in a square lattice with a pitch
of $100\,\mu$m. Both the etching and the sputtering are standard
processes that can be accurately controlled to give reproducible
results and to make large numbers of mirors in a single batch. In
the rest of this paper we analyse and measure directly how the
pyramids respond to polarised and unpolarised light.  We also test
a macroscopic model to show that this silicon pyramid mirror
geometry is suitable for making a MOT.

\begin{figure}
\includegraphics[width=8.5cm]{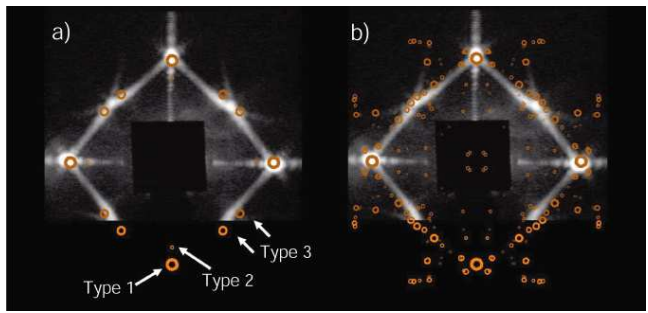}\\
\caption{Measured intensity distribution of reflected light, at a
distance of $7\,$cm from the array of pyramids, when it is
illuminated at normal incidence. A central bright spot, which is
caused by reflection from the region between pyramids, was blocked
to improve the visibility of light reflected from the pyramids.
The circles in (a) show the reflection pattern expected for a
perfect pyramid, while the circles in (b) indicate the calculated
reflection pattern for a pyramid with rounded corners. Size
indicates expected relative intensity. }\label{reflectedIntensity}
\end{figure}

\begin{figure}
\includegraphics[width=8.5cm]{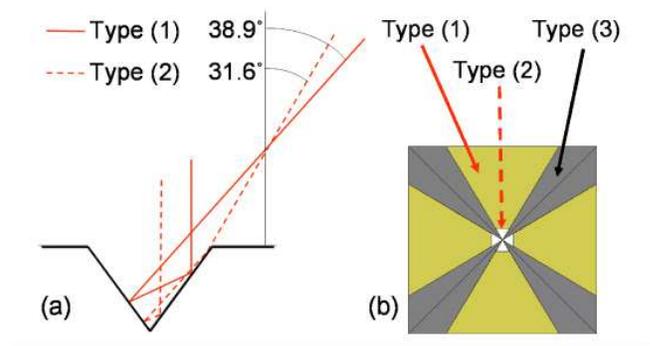}\\
\caption{(a) Cross section in the $x-z$ plane through a pyramid,
showing type (1) and type (2) trajectories.  These involve
reflections from mirrors on opposite sides of the pyramid. (b)
View of the entrance aperture of the pyramid, showing the regions
that produce type(1), type (2) and type (3) rays. } \label{rays}
\end{figure}

 The sides of the pyramids define $x$ and $y$ axes, as shown in
Fig.\,\ref{SEM}. Our first test of the mirrors is to illuminate
them with a collimated $1\,$mm-diameter laser beam (wavelength
$633\,$nm) propagating along the $z$ axis, i.e. normal to the
silicon surface and along the symmetry axis of the pyramids.
Fig.\,\ref{reflectedIntensity}\,(a) shows the reflected pattern of
light observed on a screen $7\,$cm away from the mirrors. On this
image we have drawn circles indicating the position of spots as
expected from a perfect pyramid. The three prominent spots at the
corners of the square are due to doubly reflected rays, which we
classify as type (1). These reflect from opposite faces of the
pyramid, as illustrated by the solid line in
Fig.\,\ref{rays}\,(a). There should be a fourth spot at the bottom
of the photographs, but this is blocked by a mount holding the
beamsplitter through which the array is illuminated.

 If the angle between opposite mirrors is
$\alpha$, the type (1) beams make an angle of $(\pi - 2 \alpha)$
with the $z$ axis. From the angles measured, we find that
$\alpha=(70.6\pm 0.7)^\circ$, in agreement with the expected angle
between opposing faces of $\arccos(1/3)=70.5^\circ$.

When the incident ray is close to the
apex of the pyramid (within $1.6\,\mu$m for a pyramid of
$30\,\mu$m base length), it is reflected twice by the first
mirror, as illustrated by the dashed line in Fig.\,\ref{rays}\,(a).
These rays, which we call type (2), should produce secondary spots
just inside the type (1) spots. However, the power in the type (2)
reflected beams is expected to be 100 times smaller because of the
small area from which they originate, as shown in
Fig.\,\ref{rays}\,(b).
Consequently it is not possible to identify the type (2)
beams clearly against the diffracted wings of the type
(1) beams.  Furthermore, there is a background of light
along the $x$- and $y$-axes caused by reflection from
rounded edges on the entrance aperture of the pyramid,
which can be seen in Fig.\,\ref{SEM}\,(b).

\begin{figure}[!htb]
\includegraphics[width=8.5cm]{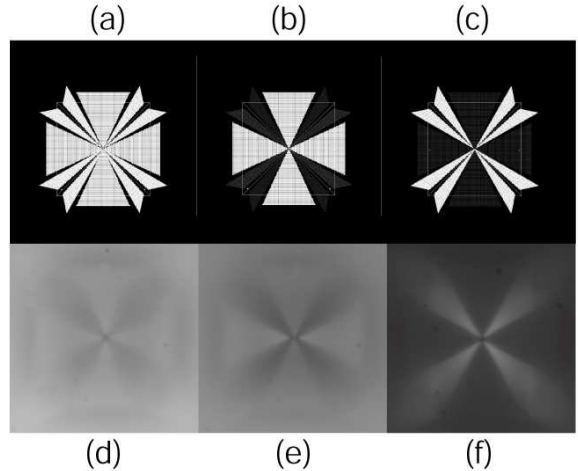}\\
\caption{Views of the vertex pyramidal mirror under an optical
microscope at $100\times$ magnification.  Top row: Raytracing
simulation. Bottom Row: Photographs. (a) and (d): without
polarisers; (b) and (e): parallel polariser and analyser; (c) and
(f): crossed polariser and analyser.} \label{microscopePics}
\end{figure}

If a ray is incident near one of the corners of the pyramid, the
first reflection sends it off towards the opposite mirror, but it
is intercepted and deflected by the adjacent mirror before the
opposite mirror sends it out of the pyramid as a type (3) ray.
These rays make an angle of $31.5^\circ$ with the $z$ axis and
form double spots at azimuthal angles of $36.9^\circ, 53.1^\circ$,
etc. as shown in Fig.\,\ref{reflectedIntensity}\,(a). These spots
are less distinct than those of type (1) because the corners of
the pyramid are rounded, a feature that does not affect the type
(1) rays. Fig.\,\ref{reflectedIntensity}\,(b) shows the same
photographed reflection pattern, but here the superimposed circles
indicate the expected position and magnitude of spots reflected
from a pyramid with rounded corners. The roundness is included in
the ray-tracing model by four additional surfaces at each corner.
These are shaped to form approximate cone sections with radii of
$2.5\,\mu$m at the base and $0.825\,\mu$m at the apex of the
pyramid. The resulting reflection pattern closely matches the
photographed intensity distribution.

The three types of ray described above also present different
characteristics when observed using polarised light. Type (1)
reflections leave the linear polarisation of the light unchanged,
whereas the type (3) reflections produce rotations of
$\pm53^\circ$ or $\pm78^\circ$. This is investigated in our second
test of the mirrors, in which we examine them with white light
under an optical microscope, illuminating them  once again along
the $z$ axis. Fig.\,\ref{microscopePics}\,(a) shows the image
calculated by ray-tracing for unpolarised light with the
microscope focussed in the plane of the apex of a perfect pyramid.
In this figure most of the area is bright. In
fig.\,\ref{microscopePics}\,(b) we show the expected image for
linearly polarised light, viewed through a parallel analyser,
which suppresses the type (3) contribution. This leads to a
reduction in the intensity of reflections from the corner region.
In Fig.\,\ref{microscopePics}\,(c), the analyser is crossed with
the polariser and only type (3) rays contribute, making the corner
region bright. The intensity patterns observed in the laboratory
are shown in figures \,\ref{microscopePics}\,(d),
\,\ref{microscopePics}\,(e), and \,\ref{microscopePics}\,(f). They
correspond closely to the calculated distributions, indicating
that the pyramid reflects light as expected.

\begin{figure}
\includegraphics[width=8.5cm]{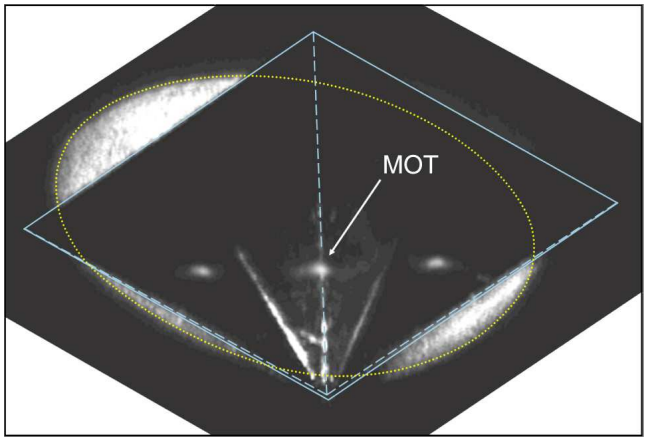}\\
\caption{Fluorescence image of $1.6\times10^8$ atoms
magneto-optically trapped in a $70.5^\circ$ pyramid. Two
reflections are also visible.  The $16.3\,$mm aperture of the
pyramid (solid line) and its sloping edges (dashed lines) have
been added to guide the eye of the reader. The outline of the
incident laser beam is shown dotted. Some light is scattered from
the edges of the pyramid and from the plane outside.}
\label{motpic}
\end{figure}

Our immediate application for these structures is to build an
array of small magneto-optical traps, integrated into an atom chip
in a single additional etching step. As in a $90^\circ$ pyramid
MOT\cite{pyrMOT}, lateral confinement is given by the first
reflections of the type-1 beams, while the vertical trapping
forces arise from the input beam and the second reflections of the
type-1 beams. In the present pyramid the beams are not orthogonal
and there are additional rays, which could also disturb the
balance of forces in the trap. To test whether it is nonetheless
possible to trap atoms in such a pyramid, we constructed a
macroscopic glass model with a base length of $16.3\,$mm, coated
with aluminium and a protective layer of Si0$_2$.
Fig.\,\ref{motpic} shows the fluorescence image from a cloud of
atoms trapped in this pyramid MOT, together with two reflections
of the cloud. There are also reflections from the top face of the
pyramid and from the edges. In order to assist the eye, we
superimpose the entrance aperture of the pyramid as a solid line
and we show the edges dashed.  We have also built and tested a
$90^\circ$ pyramid MOT of similar volume and with the same
coating, and find that there is no significant difference in the
number of trapped atoms or in the stability of the MOT.

In the microscopic version, we anticipate using pyramids with a
$200\,\mu$m base and will supply the required magnetic quadrupole
field using existing microfabrication
methods\cite{{chipReview},{MO}} to produce small current loops
around each pyramid. We estimate that such a MOT can collect as
many as 1000 atoms or as few as 1, according to the choice of
operating parameters. Compared with other methods of creating
arrays of microscopic traps on a
chip\cite{{wireGridMOTs},{edVideoMOTs}}, this relies on a simple
fabrication method and requires only a single input laser beam to
give all the necessary trapping beams. It has been shown that
Bose-Einstein condensation can be achieved on atom chips, both
with current-carrying wire traps\cite{firstChipBEC} and with
permanent magnet traps \cite{videoBEC}. Consequently it may be
possible to create an array of condensates loaded from these MOTs.
Alternatively, if there is just one atom per site, the array would
have possible applications in quantum information
processing\cite{ScheelQIP}.

Further potential applications for the pyramids are in the areas
of photonics and telecommunications. For example, by filling the
pits with ferroelectric material or liquid crystals and applying
an electric field, it may be possible to use the pyramids as fast
optical switches.

In summary, we have designed, fabricated and characterized a new
type of micro-mirror, produced by anisotropic etching through
square apertures on a silicon single crystal. As an elementary
component for optics, the micro-mirror has a variety of possible
applications in MOEMS devices. We have demonstrated that it is
possible to form a magneto-optical trap with this mirror geometry,
making these pyramids very promising for creating arrays of
microscopic traps on atom chips. Detailed experiments and further
theoretical
analysis are currently under way to develop these applications.\\

\begin{acknowledgments}
We acknowledge support from the UK EPSRC Basic Technology, QIPIRC,
and Physics programmes and from the FASTNET and Atom Chips
networks of the European Commission.
\end{acknowledgments}


\end{document}